# AC Current-Driven Magnetization Switching and Nonlinear Hall Rectification in a Magnetic Topological Insulator

Yuto Kiyonaga[1][†], Masataka Mogi[1][†]*, Ryutaro Yoshimi[2,3], Yukako Fujishiro[2,4], Yuri Suzuki[1], Max T. Birch[2], Atsushi Tsukazaki[1], Minoru Kawamura[2], Masashi Kawasaki[1,2] & Yoshinori Tokura[1,2,5]

[1]Department of Applied Physics and Quantum-Phase Electronics Center (QPEC), University of Tokyo, Bunkyo-ku, Tokyo, Japan

[2]RIKEN Center for Emergent Matter Science (CEMS), Wako, Saitama, Japan

[3]Department of Advanced Materials Science, University of Tokyo, Kashiwa, Chiba, Japan

[4]RIKEN Cluster for Pioneering Research (CPR), Wako, Saitama, Japan

[5]Tokyo College, University of Tokyo, Bunkyo-ku, Tokyo, Japan

[†]These authors contributed equally: Yuto Kiyonaga, Masataka Mogi.

*e-mail: mogi@ap.t.u-tokyo.ac.jp

**Abstract:** Spin-orbit torque arising from the spin-orbit-coupled surface states of topological insulators enables current-induced control of magnetization with high efficiency. Here, alternating-current (AC) driven magnetization reversal is demonstrated in a semi-magnetic topological insulator $(Cr,Bi,Sb)_2Te_3/(Bi,Sb)_2Te_3$, facilitated by a low threshold current density of $1.5 \times 10^9 \, Am^{-2}$. Time-domain Hall voltage measurements using an oscilloscope reveal a strongly nonlinear and nonreciprocal Hall response during the magnetization reversal process. Fourier analysis of the time-varying Hall voltage identifies higher-harmonic signals and a rectified direct-current (DC) component, highlighting the complex interplay among the applied current, external magnetic field, and magnetization dynamics. Furthermore, a hysteretic behavior in the current-voltage characteristics gives rise to frequency mixing under dual-frequency excitation. This effect, distinct from conventional polynomial-based nonlinearities, allows for selective extraction of specific frequency components. The results demonstrate that AC excitation can not only switch magnetization efficiently but also induce tunable nonlinear responses, offering a new pathway for multifunctional spintronic devices with potential applications in energy-efficient memory, signal processing, and frequency conversion.

## 1. Introduction

The interplay of electron spin and charge transport underlies a wide range of spintronic phenomena and applications, including current-induced magnetization reversal in ferromagnets via spin-transfer and spin-orbit torques.[1,2,3] Beyond acting as the driving force for magnetization dynamics, the dynamics of localized spins significantly influence the charge transport properties, leading to nonreciprocal conduction,[4,5] spin-motive forces,[6,7] and emergent electric fields.[8,9,10] These effects open possibilities for advanced functionalities, such as diode operation, DC rectification, frequency conversion, and novel inductive elements.[10,11]

Nonlinear Hall effects arising from such spin-charge coupling also offer a means to detect current-driven magnetization dynamics.[12,13] In particular, second harmonic Hall voltages generated by continuous AC current excitation are commonly employed to probe magnetization oscillations.[13] However, driving magnetization dynamics typically requires large current densities on order of $10^{10}$ to $10^{11}$ $Am^{-2}$, often applied as short pulses to mitigate Joule heating,[1] posing multiple challenges originating from extrinsic thermal effects, including parasitic Nernst effects,[14] spin-Seebeck effects,[15] and enhanced magnon scattering.[4,5] These parasitic effects often mask the intrinsic nonlinear signals that stem directly from magnetization dynamics. To date, a nonlinear Hall effect directly associated with magnetization reversal driven by spin-transfer or spin-orbit torques remains largely unexplored.

Time-domain measurements of nonlinear Hall effect provide an effective method for investigating magnetization dynamics. Such measurements have been employed to study ultrafast magnetization switching,[16,17] magnetic domain wall motion induced by pulse current,[18] and inertial skyrmion dynamics under AC currents.[19] In our study, by simultaneously monitoring the longitudinal resistance as a thermometer, thermal effects can be evaluated, helping to distinguish intrinsic nonlinear signals from parasitic heating effects. Moreover, real-time current-Hall voltage characteristics allow direct analysis of nonlinear behavior and phase, making time-domain measurements well suited for exploring nonlinear Hall effects in AC-driven magnetization dynamics.

Topological insulators (TIs) serve as a unique platform to explore these effects, as they host insulating bulk states and conducting surface states with strong spin-orbit coupling, where electron spin is locked perpendicularly to its momentum.[20] This property facilitates efficient spin-charge conversion via spin-orbit torques (SOTs).[21] Indeed, in heterostructures composed of TIs and various ferromagnets, current-driven magnetization switching has been demonstrated at low current densities on the order of $10^9$ to $10^{10}$ Am$^{-2}$.[13,22,23] Furthermore, when magnetism is introduced into the topological surface state, a magnetization gap is induced, leading to a large anomalous Hall conductivity due to Berry curvature.[24,25] Nonetheless, as in conventional ferromagnets, nonlinear transport arising from magnetization dynamics can still be masked by substantial magnon scattering under low current excitation.[4,5]

In this study, we demonstrate AC current-induced magnetization dynamics and its associated nonlinear Hall effect in a semi-magnetic TI (Cr,Bi,Sb)$_2$Te$_3$/(Bi,Sb)$_2$Te$_3$ heterostructure thin film (see Methods for details). We directly observe magnetization reversal in response to AC current drive by measuring the Hall voltage in time domain using an oscilloscope, The measured Hall voltage exhibits strong current nonlinearity and higher-harmonic signals, as well as a rectified DC offset. A systematic analysis of the current-Hall voltage characteristics, combined with Fourier transforms, reveals notable nonlinear and hysteretic responses in the Hall voltage, shedding light on the interplay between current, external magnetic field, and magnetization dynamics. We also identify an asymmetric frequency-mixing phenomenon originating from the magnetization switching hysteresis, which contrasts with conventional polynomial-type nonlinearities.

## 2. AC current-induced magnetization switching

For current-induced perpendicular magnetization switching in TIs, the damping-like SOT $\vec{\sigma} \times (\vec{\sigma} \times \vec{M})$ plays a central role,[3] where $\sigma$ is the spin polarization of conduction electron and $M$ is the localized magnetization. In TIs, the spins are polarized perpendicular to the current when the chemical potential lies in the surface state.[24] Therefore, the magnetization with perpendicular magnetic anisotropy is tilted toward the current direction[5,13,23] (**Figure 1a**). The resulting magnetization can be detected by the anomalous Hall effect (AHE) arising from the magnetization gap in the TI surface state (**Figure 1b**).

We first confirm current-induced magnetization switching by applying high-current pulses (up to $I_{pulse} = \pm 500$ μA, or current density of $\pm 5 \times 10^9$ Am$^{-2}$), and under an in-plane magnetic field $B_x = 0.01$ T, followed by measuring the Hall voltage at a low sensing current ($I_x = 1$ μA, or current density of $1 \times 10^7$ Am$^{-2}$) (**Figure 1c**). The Hall voltage $V_y$ is given by $V_y = R_{yx}I_x \propto M_z I_x$, where $M_z$ is the out-of-plane component contributing to the AHE. Hence, the sign of the Hall resistance $R_{yx}$ directly reflects the direction of $M$. By varying the amplitude of the pulse, we observe a clear sign reversal of $R_{yx}$ at a threshold current of about 150 µA (current density of $1.5 \times 10^9$ Am$^{-2}$) (Figure 1e). The switching polarity reverses if we flip either the current direction or the in-plane field direction, consistent with SOT-driven perpendicular-magnetization switching.

Next, instead of applying current pulses, we record the Hall voltage in real time with an oscilloscope (see Methods), while applying an AC current excitation $I_x = I_{peak} \sin(2\pi f t)$ with a peak current amplitude of $I_{peak} = 300$ µA at a frequency of $f = 101$ Hz (**Figure 1d**). We evaluate $R_{yx}$ from the $R_{yx} = V_y/I_x$ and plot as a function of $I_x$ in Figure 1f. When $0$ µA $< I_x <$ 300 µA, the Hall resistance $R_{yx}$ gradually decreases and switches polarity above a threshold current (defined where it crosses $R_{yx} = 0$) of about 150 µA. Once the polarity is switched, it remains until it is switched again at around $I_x = -150$ µA. This is consistent with the case of pulse-current-induced magnetization reversal discussed above. Simultaneous monitoring of the longitudinal resistance $R_{xx}$ indicates that the temperature is kept below 30 K during the measurement, which is lower than $T_C \sim 40$ K and there is no temperature hysteresis (see Supplementary Note 3), confirming that heating is not a primary origin of the observed magnetization reversal.

### 3. Nonlinear Hall voltage and hysteretic behavior

**Figure 2a** shows the time-dependent evolution of the Hall voltage $V_y$(t). Since the Hall resistance $R_{yx}$ becomes negative for $I_x > 0$ and positive for $I_x < 0$, as shown in **Figure 2b**, we see that $V_y$ takes predominantly negative values, reflecting the reversed magnetization state when $I_x$ is above the threshold. Meanwhile, small positive spikes (black triangles in Figure 2a) appear, indicating the short period when the magnetization remains unreversed below the threshold current. To clarify the role of the current amplitude, we compare waveforms of the Hall voltage for several peak current values (**Figure 2c**). At a low peak current value ($I_{peak} =$ 14 µA) which is far below the threshold current, Hall voltage shows a nearly sinusoidal time dependence with alternating signs, originating from the out-of-plane component of magnetization under $B_x \sim 0.01$ T. With increasing $I_{peak}$ (from 71 to 300 µA), the waveforms become increasingly distorted, revealing pronounced nonlinearity in the Hall response. Figure 2d further highlights this nonlinearity by plotting the $I_x - V_y$ relationship for each $I_{peak}$. At $I_{peak} = 14$ µA, a nearly linear relationship appears, while at higher currents (e.g., 300 µA), the curves develop a butterfly-shaped hysteresis: the Hall voltages differ for increasing vs decreasing current. Such a hysteresis results from the magnetization dynamics and phase delay when the applied AC current exceeds the threshold of magnetization switching, representing continuous magnetization reversal.

The nonlinear Hall response is also strongly affected by the in-plane magnetic field altering the magnetization orientation. To investigate this, we perform time-domain measurements of $V_y$ at a larger field ($B_x = 2$ T). Figure 3a shows the contrastive Hall voltage waveforms at $B_x = 0.01$ T and 2 T (the former being the same data from Figure 2a). Unlike at 0.01 T, at 2 T, which is strong enough to force the magnetization to lie in-plane, no positive spikes appear in regions i and iii of **Figure 3a**, and the signal remains consistently negative. Correspondingly, the $I_x - V_y$ characteristic shows rectifying behavior without hysteresis (**Figure 3b**, bottom panel). A plausible interpretation is that, under the strong in-plane field, the magnetization never switches. Instead, magnetization acquires a finite $M_z$ component in the $-z$ direction for $I_x > 0$ and in the $+z$ direction for $I_x < 0$, resulting in a rectified Hall response.

We note that extrinsic thermoelectric effects[14,15,26] as well as asymmetric scattering from magnon emission/absorption[4,5] may contribute to the nonlinear Hall signals discussed above, complicating the quantitative separation of each effect. However, none of these effects can cause the hysteresis observed at $B_x = 0.01$ T. Therefore, the main contribution of the nonlinear signals originates from magnetization dynamics, particularly in the case of $B_x = 0.01$ T.

To systematically compare these nonlinear signals, we decompose the time-domain Hall voltage $V_y(t)$ via Fourier transforms. In a simple power-series expansion of $V_y$ in terms of current $I_x(t) = I_0 \sin(\omega t)$, $V_y(t)$ can be expressed as:

$$V_y(t) = V_y^{(0)} + V'^{(1)}_y \sin(\omega t) + V'^{(2)}_y \cos(2\omega t) + V'^{(3)}_y \sin(3\omega t) + V'^{(4)}_y \cos(4\omega t) + \cdots,$$

where $V_y^{(0)}$ and $V'^{(n)}_y$ ($n = 1, 2, 3, 4, \cdots$) are a constant and coefficients, respectively (see Supplementary Note 4 for details). Here all the odd-harmonic components appear as sine functions, while all the even-harmonic components appear as cosine functions in this power series. More generally, a phase delay arising from hysteresis is described by adding odd-harmonic cosine functions and even-harmonic sine functions as,

$$V_y(t) = V_y^{(0)} + V'^{(1)}_y \sin(\omega t) + V''^{(1)}_y \cos(\omega t) + V'^{(2)}_y \cos(2\omega t) + V''^{(2)}_y \sin(2\omega t)$$
$$+ V'^{(3)}_y \sin(3\omega t) + V''^{(3)}_y \cos(3\omega t) + V'^{(4)}_y \cos(4\omega t) + V''^{(4)}_y \sin(4\omega t) + \cdots,$$

where $V''^{(n)}_y$ ($n = 1, 2, 3, 4, \cdots$) are coefficients of these phase-shifted components. **Figure 3c** shows the Fourier amplitudes at $B_x = 0.01$ T and 2 T. In this figure, gray, red, and blue bars represent the components $V^{(0)}_y$, $V'^{(n)}_y$, and $V''^{(n)}_y$, respectively. Both cases exhibit strong second-harmonic $V'^{(2)}_y$ appear, corresponding to the fact that $V_y$ remains negative for both $I_x > 0$ and $I_x < 0$. However, additional phase-shifted components represented as $V''^{(n)}_y$ appear only at $B_x = 0.01$ T, reflecting the delayed, hysteretic response of magnetization reversal. In contrast, at 2 T, the magnetization follows $I_x$ more smoothly, eliminating large phase shifts.

## 4. Asymmetric frequency mixing

Finally, we demonstrate a frequency-mixing phenomenon[27] enabled by AC current-induced magnetization switching. In general, a nonlinear system driven by two frequencies $f_1$ and $f_2$ can generate signals at their sum $f_1 + f_2$ and difference $|f_1 - f_2|$. If the nonlinearity is purely polynomial in current, the amplitudes of the $f_1 + f_2$ and $|f_1 - f_2|$ components are expected to be equal[27] (See Supplementary Note 5). However, in our semi-magnetic TI, the hysteresis of magnetization reversal breaks this symmetry. In the experiment, we apply $I_x(t) = I_1\sin(2\pi f_1 t) + I_2\sin(2\pi f_2 t)$ with $I_1 = I_2 = 150$ μA, $f_1 = 37$ Hz, and $f_2 = 125$ Hz. When only a single frequency ($f_1 = 37$ Hz or $f_2 = 125$ Hz) is applied (peak current 300 μA), the response is similar to our earlier single-frequency result (**Figure 4a**), confirming that the hysteretic behavior is governed by magnetization switching (Figure 1e,f), rather than the inertia dynamics of magnetic domains or capacitive/inductive components of the electrical circuits. With both frequencies present (**Figure 4b**), we observe broadband wave mixing, ranging from DC component to 338 Hz in the Fourier spectrum of $V_y$ (**Figure 4c**). Notably, at a low magnetic field (0.01 T), the $f_1 + f_2$ component is substantially larger than the $|f_1 - f_2|$ component. In contrast, at a high field (2 T), both peaks exhibit similar amplitudes, indicating a more conventional polynomial-type nonlinearity. A numerical simulation (**Figure 4d**), assuming a simplified $I_x - V_y$ characteristic with a well-defined threshold (insets of Figure 4d), reproduces this asymmetry only when a finite threshold current $I_{th} = 150$ μA is considered (mimicking the low-field case). If $I_{th} = 0$ μA (high-field case), the two sidebands remain equal. Hence, the hysteretic magnetization reversal leads to asymmetric frequency mixing.

## 5. Conclusion

Our findings demonstrate that AC current can induce continuous magnetization reversal in a semi-magnetic TI heterostructure, at a remarkably low current density ($1.5 \times 10^9$ Am$^{-2}$). We have clarified the nonlinear Hall effect accompanying this process and shown that the hysteretic and phase-delayed Hall responses are governed by a threshold current for magnetization reversal. Furthermore, the strength of the magnetic field can control the presence or absence of the hysteresis. The pronounced nonlinear Hall effect observed here holds promise for Hall rectification,[28,29] which has recently been studied for terahertz-to-DC conversion[30] or AC-to-DC conversion.[31] Additionally, when a current containing components of two different frequencies is applied, we discover a distinctive frequency-mixing effect where the magnitudes of the sum-frequency and difference-frequency components differ due to the hysteresis for magnetization reversal. Such frequency-mixing effects are commonly used for technologies such as photoacoustic imaging[32] and microwave generation[33]. The electrical asymmetric frequency mixing effect induced by the nonlinearity of magnetization reversal process may thus be leveraged in spintronic devices for selective extraction of desired frequency components. Furthermore, in the field of neural networks, elements that combine nonlinearity with short-term memory are widely used for physical reservoir computing.[32-37] The nonlinear and hysteretic response discussed here also has the potential for nonlinear transformation of inputs into higher-dimensional outputs, making our system promising as a reservoir. While all the responses explored in this study are at 2.5 K, the use of ferromagnetic materials with high transition temperatures in the magnetic layer exchange-coupled to the topological insulator could lead to potential functionality even at room temperature. [40,41,42] Thus, our study paves the way for harnessing hysteretic magnetization dynamics, with potential applications in spin-orbit-based low-power switching elements, and advanced nonlinear electronics such as neuromorphic computing.[43]

## 6. Experimental Section/Methods

*Sample fabrication and electric transport measurement*: We grew $(Cr,Bi,Sb)_2Te_3/(Bi,Sb)_2Te_3$ thin films on InP(111) substrates by molecular beam epitaxy under the base pressure on the order of $10^{-7}$ Pa. The flux ratio $P_{Cr}:P_{Bi}:P_{Sb}:P_{Te} = 1:9:16:1000$ was used to tune the Fermi energy inside the bulk gap. The films are fabricated into 10 μm-wide Hall-bar devices. The structures of them are illustrated in Supplementary Figure S1(a). In a typical sample, we measured the electrical transport properties using Physical Property Measurement System (PPMS), as shown in Supplementary Figure S1(b). The longitudinal and Hall resistance of a typical sample is about 10 kΩ and 2 kΩ, respectively. The Hall conductivity at the lowest temperature $T = 2.5$ K is about $0.3 \times e^2/h$, which indicates the Fermi level is fairly close to the magnetization gap. From the temperature dependence of Hall and longitudinal resistivity, the Curie temperature can be determined as $T_C \sim 40$ K.

*Time-domain measurement*: We measured the Hall voltage as well as the longitudinal voltage in real time with an oscilloscope and a current source (Keithley 6221). Set up in a PPMS chamber is shown in Supplementary Figure S2. The oscilloscope monitors the Hall voltage $V_y(t)$ as

$$V_y(t) = V_{CH2}(t) - V_{CH1}(t)$$

and the voltage on the resistor $V_R$ as

$$V_R(t) = V_{CH3}(t)$$

where $V_{CH1}(t)$, $V_{CH2}(t)$, and $V_{CH3}(t)$ are the measured voltages of the channels CH1, CH2, and CH3, respectively. We note that the current flowing through the circuit is obtained by

$I_x(t) = \frac{V_R(t) \, [V]}{1000 \, [\Omega]} = \frac{V_{CH3}(t) \, [V]}{1000 \, [\Omega]}$.

**Acknowledgements:** We thank Tomoyuki Yokouchi and Lixuan Tai for stimulating discussions. This work was supported by JSPS KAKENHI Grant Nos. JP22H04958, JP23H05431 and JP24K16986, and JST PRESTO Grant No. JPMJPR23HA.

**Author contributions:** M.M. and Y.T. conceived the study. Y.K. and M.M. grew the samples with help of R.Y., K.S.T., A.T. and M. Kawas. Y.K. and M.M. performed measurements with help of Y.F., Y.S., M.T.B. and M.Kawam. All authors discussed the results. Y.K., M.M. and Y.T. wrote the manuscript with inputs from all other authors. Y.T. supervised the project.

**References:**

[1] Liu, L. et al. Current-Induced Switching of Perpendicularly Magnetized Magnetic Layers Using Spin Torque from the Spin Hall Effect. *Phys. Rev. Lett.* **109**, 096602 (2012)

[2] Ralph, D. C. & Stiles, M. D. Spin transfer torques. *J. Magn. Magn. Mater.* **320**, 1190–1216 (2008)

[3] Manchon, A. et al. Current-induced spin-orbit torques in ferromagnetic and antiferromagnetic systems. *Rev. Mod. Phys.* **91**, 035004 (2019)

[4] Yasuda, K. et al. Large Unidirectional Magnetoresistance in a Magnetic Topological Insulator. *Phys. Rev. Lett.* **117**, 127202 (2016)

[5] Yasuda, K. et al. Current-Nonlinear Hall Effect and Spin-Orbit Torque Magnetization Switching in a Magnetic Topological Insulator. *Phys. Rev. Lett.* **119**, 137204 (2017)

[6] Yang, S. A. et al. Universal Electromotive Force Induced by Domain Wall Motion. *Phys. Rev. Lett.* **102**, 067201 (2009)

[7] Emori, S. et al. Current-driven dynamics of chiral ferromagnetic domain walls. *Nat. Mater.* **12**, 611–616 (2013)

[8] Nagaosa, N. & Tokura, Y. Emergent electromagnetism in solids. *Phys. Scr.* **2012**, 014020 (2012)

[9] Schulz, T. et al. Emergent electrodynamics of skyrmions in a chiral magnet. *Nat. Phys.* **8**, 301–304 (2012)

[10] Yokouchi, T. et al. Emergent electromagnetic induction in a helical-spin magnet. *Nature* **586**, 232–236 (2020)

[11] Yamane, Y., Fukami, S., & Ieda, J. Theory of Emergent Inductance with Spin-Orbit Coupling Effects. *Phys. Rev. Lett.* **128**, 147201 (2022)

[12] Sala, G. et al. Real-time Hall-effect detection of current-induced magnetization dynamics in ferrimagnets. *Nat. Commun.* **12**, 656 (2021)

[13] Fan, Y. et al. Magnetization switching through giant spin–orbit torque in a magnetically doped topological insulator heterostructure. *Nat. Mater.* **13**, 699–704 (2014)

[14] Avci, C. O. et al. Interplay of spin-orbit torque and thermoelectric effects in ferromagnet/normal-metal bilayers. *Phys. Rev. B* **90**, 224427 (2014)

[15] Uchida, K. et al. Observation of the spin Seebeck effect. *Nature* **455**, 778–781 (2008)

[16] Baumgartner, M. et al. Spatially and time-resolved magnetization dynamics driven by spin-orbit torques. *Nat. Nanotechnol.* **12**, 980–986 (2017)

[17] Grimaldi, E. et al. Single-shot dynamics of spin–orbit torque and spin transfer torque switching in three-terminal magnetic tunnel junctions. *Nat. Nanotechnol.* **15**, 111–117 (2020)

[18] Yoshimura, Y. et al. Soliton-like magnetic domain wall motion induced by the interfacial Dzyaloshinskii–Moriya interaction. *Nat. Phys.* **12**, 157–161 (2016)

[19] Birch, M. T. et al. Dynamic transition and Galilean relativity of current-driven skyrmions. *Nature* **633**, 554–559 (2024)

[20] Hasan, M. Z. & Kane, C. L. Colloquium: Topological insulators. *Rev. Mod. Phys.* **82**, 3045 (2010)

[21] Kondou, K. et al. Fermi-level-dependent charge-to-spin current conversion by Dirac surface states of topological insulators. *Nat. Phys.* **12**, 1027–1031 (2016)

[22] Mellnik, A. R. et al. Spin-transfer torque generated by a topological insulator. *Nature* **511**, 449–451 (2014)

[23] Mogi, M. et al. Current-induced switching of proximity-induced ferromagnetic surface states in a topological insulator. *Nat. Commun.* **12**, 1404 (2021)

[24] Tokura, Y., Yasuda, K., & Tsukazaki, A. Magnetic topological insulators. *Nat. Rev. Phys.* **1**, 126-143 (2019)

[25] Yu, R. et al. Quantized Anomalous Hall Effect in Magnetic Topological Insulators. *Science* **329**, 5987 (2010)

[26] Tai, L. et al. Giant Hall Switching by Surface-State-Mediated Spin-Orbit Torque in a Hard Ferromagnetic Topological Insulator. *Adv. Mater.* **36**, 2406772 (2024)

[27] Min, L. et al. Colossal room-temperature non-reciprocal Hall effect. *Nat. Mater.* **23**, 1671–1677 (2024)

[28] Isobe, H., Xu, S.-Y., & Fu, L. High-frequency rectification via chiral Bloch electrons. *Sci. Adv.* **6**, 13 (2020)

[29] He, P. et al. Quantum frequency doubling in the topological insulator $Bi_2Se_3$. *Nat. Commun.* **12**, 698 (2021)

[30] Zhang, Y. & Fu, L. Terahertz detection based on nonlinear Hall effect without magnetic field. *Proc. Natl. Acad. Sci. USA* **118**, e2100736118 (2021)

[31] Onishi, Y. & Fu, L. High-efficiency energy harvesting based on a nonlinear Hall rectifier. *Phys. Rev. B* **110**, 075122 (2024)

[32] Gusev, V. & Chigarev, N. Nonlinear frequency-mixing photoacoustic imaging of a crack: Theory. *J. Appl. Phys.* **107**, 124905 (2010)

[33] Rice, A. et al. Terahertz optical rectification from <110> zinc-blende crystals. *Appl. Phys. Lett.* **64**, 1324–1326 (1994)

[34] Torrejon, J. et al. Neuromorphic computing with nanoscale spintronic oscillators. *Nature* **547**, 428–431 (2017).

[35] Moon, J. et al. Temporal data classification and forecasting using a memristor-based reservoir computing system. *Nat. Electro.* **2**, 480–487 (2019)

[36] Zhong, Y. et al. Dynamic memristor-based reservoir computing for high-efficiency temporal signal processing. *Nat. Commun.* **12**, 408 (2021)

[37] Zhong, Y. et al. A memristor-based analogue reservoir computing system for real-time and power-efficient signal processing. *Nat. Electro.* **5**, 672–681 (2022)

[38] Yokouchi, T. et al. Pattern recognition with neuromorphic computing using magnetic field–induced dynamics of skyrmions. *Sci. Adv.* **8**, 39 (2022)

[39] Liang, X. et al. Physical reservoir computing with emerging electronics. *Nat. Electro.* **7**, 193–206 (2024)

[40] Wang, Y. et al. Room temperature magnetization switching in topological insulator-ferromagnet heterostructures by spin-orbit torques. *Nat. Commun.* **8**, 1364 (2017)

[41] Wang, H. et al. Room temperature energy-efficient spin-orbit torque switching in two-dimensional van der Waals $Fe_3GeTe_2$ induced by topological insulators. *Nat. Commun.* **14**, 5173 (2023)

[42] Choi, G. S. et al. Highly Efficient Room-Temperature Spin-Orbit-Torque Switching in a Van der Waals Heterostructure of Topological Insulator and Ferromagnet. *Adv. Sci.* **11**, 2400893 (2024)

[43] Liu, Y. et al. Cryogenic in-memory computing using magnetic topological insulators, Nat. Mater. DOI:10.1038/s41563-024-02088-4 (2025).

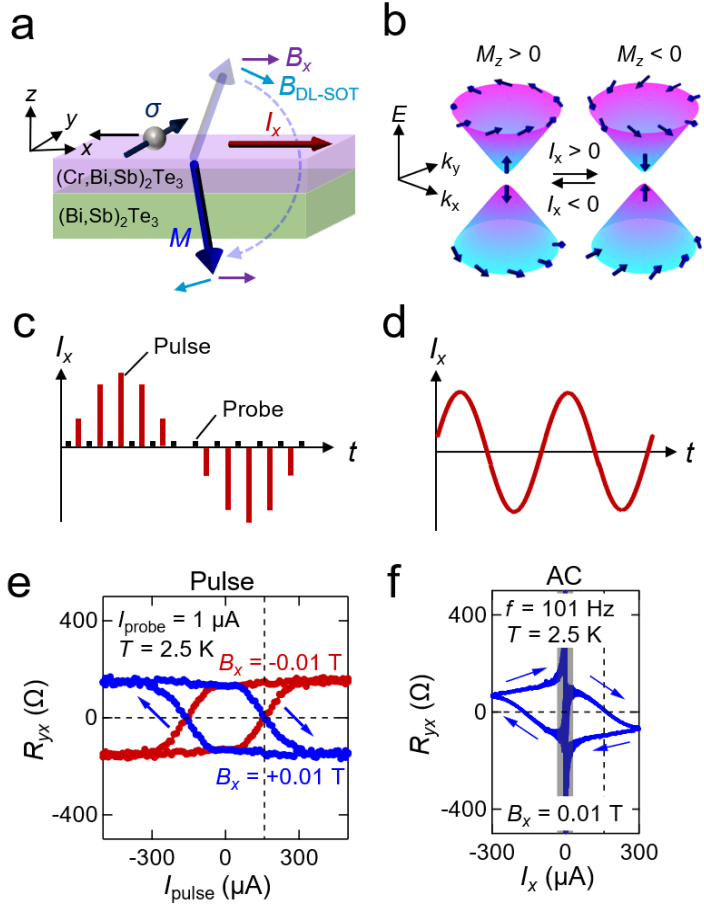

**Figure 1: Magnetization switching by pulse current and AC current. a**, Schematic illustration of current-induced magnetization switching. The purple and light blue arrows represent in-plane magnetic field $B_x$ and the effective magnetic field of damping-like spin-orbit torque $B_{\text{DL-SOT}}$, respectively. $\sigma$ denotes the spin-polarization of conduction electron. **b**, Spin-polarization of gapped Dirac surface states (blue arrows). **c,d**, Input waveform of pulse and probe current measurement **c** and alternating current measurement **d**. **e,f**, The change in Hall resistance during applying current for pulse and probe **e** and for AC **f**. For the pulse case **e**, the Hall resistance change is measured by the probe current of 1 μA under the magnetic field $B_x$ of 0.01 T (blue) and $-0.01$ T (red) applied parallel to the current direction. The transverse axis shows the pulse current value which varies from $-500$ μA to 500 μA and then from 500 μA to $-500$ μA. For the AC case **f**, the Hall resistance is measured by current $I_x(t) = I_{\text{peak}} \sin(2\pi f t)$ ($I_{\text{peak}} = 300$ μA, $f = 101\,\text{Hz}$) at each time $t$ under $B_x = 0.01$ T. The black shaded area indicates that the measured value of $R_{yx}$ diverges near $I_x = 0$ because it results in an indeterminate form (0/0). The shown data are antisymmetrized with respect to the magnetic field. The vertical dotted lines represent the threshold current where the Hall resistance changes its sign, and the blue arrows represent the direction of the flow of time.

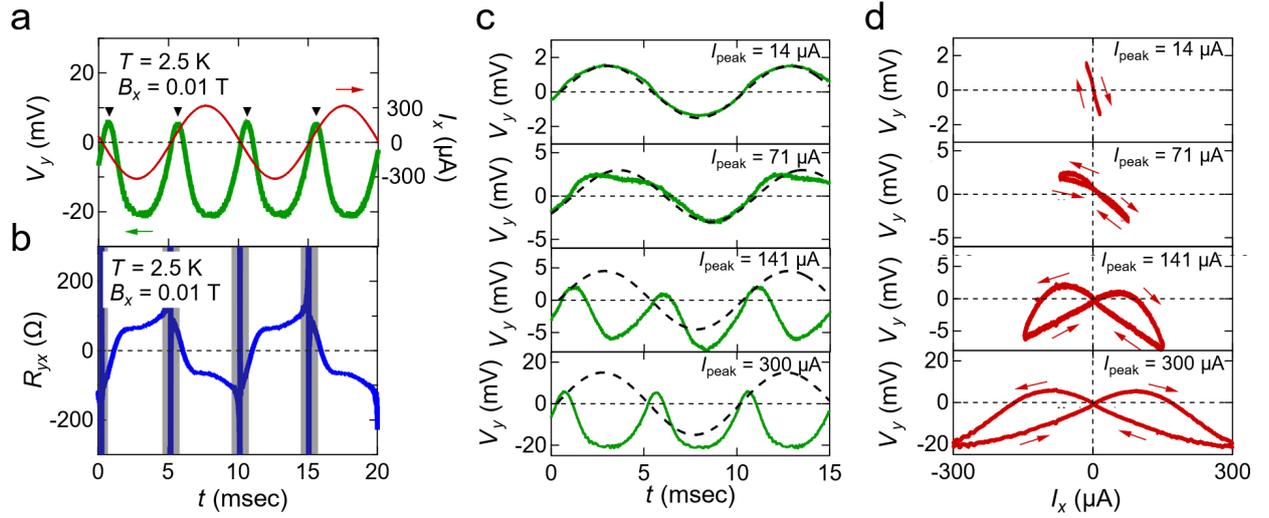

**Figure 2: Time domain measurements of magnetization switching. a,** Waveforms of input AC current (red, $f = 101\,\mathrm{Hz}$) and output Hall voltage (green) measured under in-plane magnetic field $B_x = 0.01\,\mathrm{T}$ and temperature $T = 2.5\,\mathrm{K}$. Black triangles point to positive spikes of Hall voltage. **b,** Change in Hall resistance $R_{yx}$ (blue) calculated from current and Hall voltage shown in **a**. Black shaded areas indicate that the value of $R_{yx}$ diverges around the time period around $I_x = 0$ because it results in an indeterminate form (0/0). **c,** Waveforms of the Hall voltage (green) for AC current with amplitudes of 14 μA, 71 μA, 141 μA, and 300 μA. The sinusoidal dotted curve (black) is a guide to the eye which indicates the phase of the AC current. **d,** Hall voltage vs current in the case of applying AC current whose amplitudes are 14 μA, 71 μA, 141 μA, and 300 μA. The red arrows represent the direction of the flow of time.

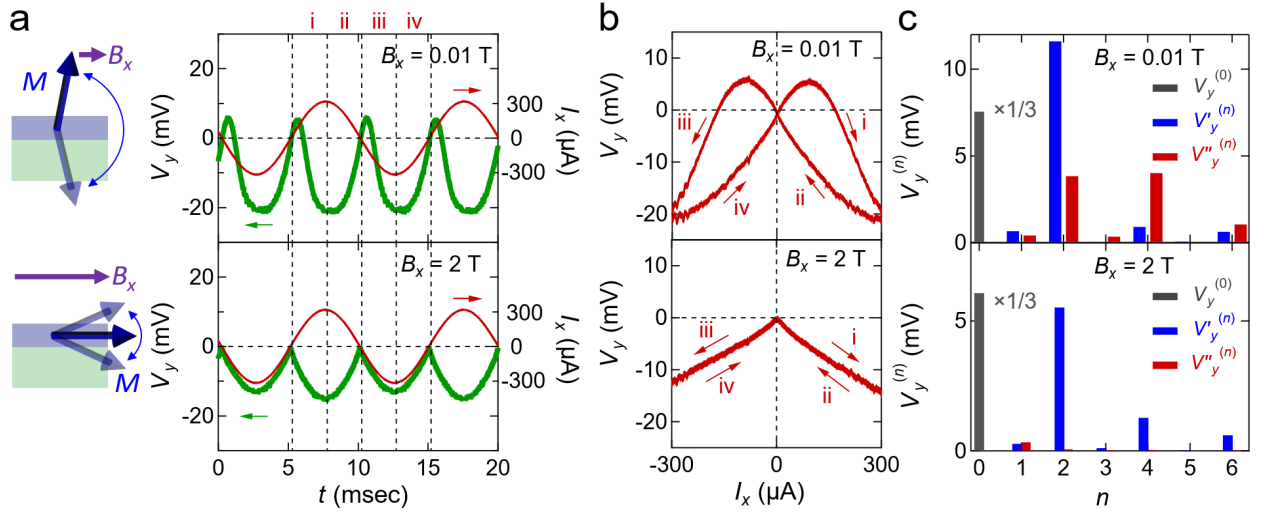

**Figure 3: Distinct nonlinear Hall responses under in-plane magnetic fields. a**, Waveform of Hall voltage (green) and current (red) under magnetic field $B_x = 0.01$ T (top) and $B_x = 2$ T (bottom) and temperature $T = 2.5$ K. The left illustrations indicate the magnetic field (purple arrows) and the magnetization when applying no current to the sample (dark blue arrows) and the magnetization oscillation (light blue arrows) under the AC current excitation. The vertical dotted lines divide one cycle into 4 regions (i: $dI_x/dt > 0$, $I_x > 0$, ii: $dI_x/dt < 0$, $I_x > 0$, iii: $dI_x/dt < 0$, $I_x < 0$, iv: $dI_x/dt > 0$, $I_x < 0$). **b**, Hall voltage vs current under magnetic field $B_x = 0.01$ T (top) and $B_x = 2$ T (bottom). The symbols i, ii, iii, and iv on the upper graph corresponds to the ones in **a**. **c**, Fourier transformation of the Hall voltage under magnetic field $B_x = 0.01$ T (top) and $B_x = 2$ T (bottom). The blue, red, and gray bars denote the in-phase components $V'^{(n)}_y$, the out-of-phase components $V''^{(n)}_y$, and the constant component $V^{(0)}_y$, respectively.

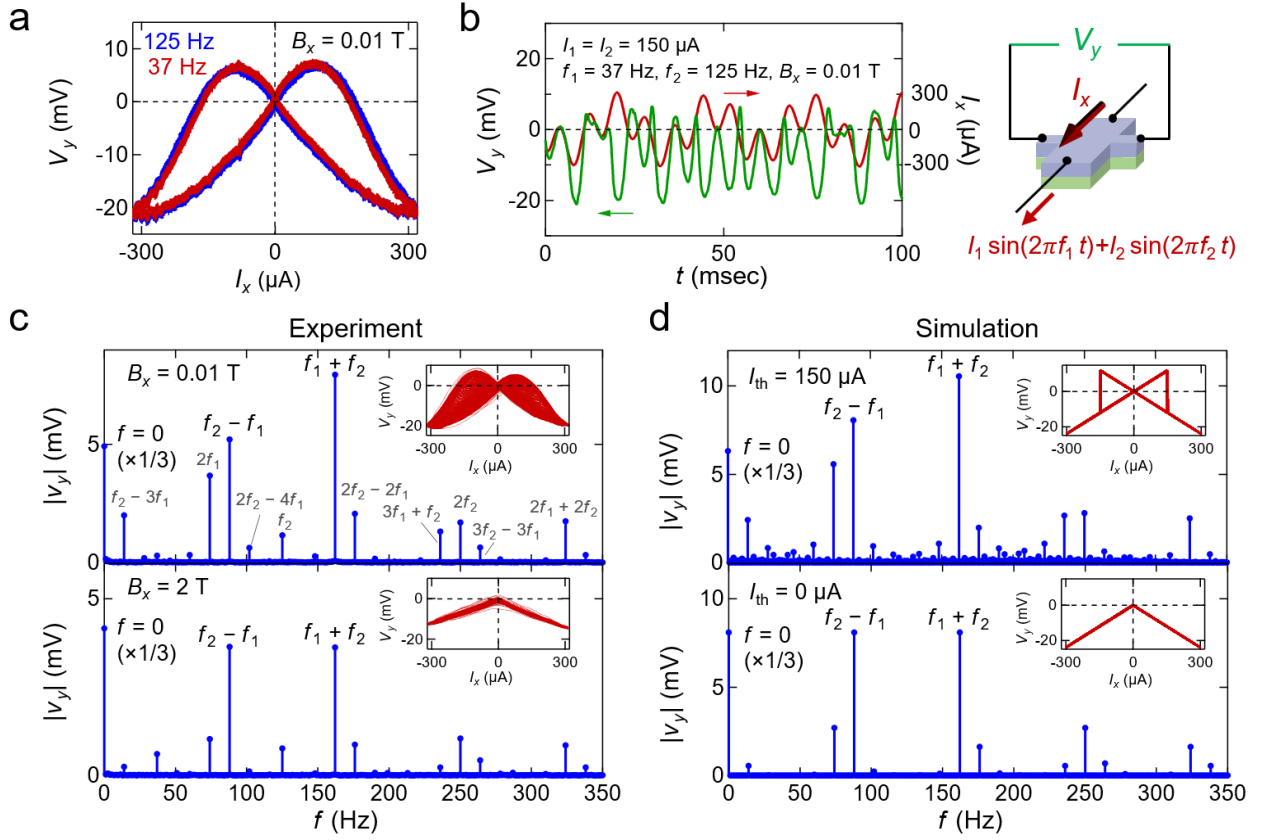

**Figure 4: Frequency-mixing accompanied by magnetization dynamics. a**, Hall voltage vs current for the AC current with the frequencies, $f_1 = 37$ Hz (red) and $f_2 = 125$ Hz (blue), and the amplitude of 300 μA under magnetic field $B_x = 0.01$ T and temperature $T = 2.5$ K. **b**, The waveform of the input current including 2 frequency components $I_x(t) = 150 \sin(2\pi f_1 t) + 150 \sin(2\pi f_2 t)$ [μA] (red) and the Hall response $V_y$ (green) under magnetic field $B_x = 0.01$ T and temperature $T = 2.5$ K. The schematic on the right side shows the experimental set-up. **c**, Absolute value of Fourier component $|v_y(f)|$ from the measured responses under magnetic field $B_x = 0.01$ T (top) and $B_x = 2$ T (bottom). The inset shows Hall voltage vs current in each case. In the all panels of **c** and **d**, the peaks of the sum-frequency $f_1 + f_2$ and the difference-frequency $|f_1 - f_2|$ is explicitly indicated. In the top panel for the experiment under 0.01 T, the peaks of other linear combinations (gray) are also indicated. **d**, Absolute value of Fourier component $|v_y(f)|$, which is defined in the same way as **c**, from the simulated responses for $I_{th} = 150$ μA (top) and $I_{th} = 0$ μA (bottom), mimicking the experimental cases for $B_x = 0.01$ T and $B_x = 2$ T, respectively. All the data in **c** and **d** are plotted as a function of $f$ with 1 Hz intervals.

**Supplementary Note 1: Analysis of the raw data in the time-domain measurements**

Using an oscilloscope, we measured Hall voltage in response to current in the time domain. To exclude the background from longitudinal voltage due to electrode misalignment, we subtracted the measured data under the magnetic field in the $-x$ direction from the one under that in the $+x$ direction (anti-symmetrization). Here we emphasize that when measuring data under the opposite field, the way of initialization is also inverted. For example, we measured $V_y(t)$ under $B_x = +0.01$ T after initializing magnetization by setting to $B_x = +2$ T, and then measured $V_y(t)$ under $B_x = -0.01$ T after setting to $B_x = -2$ T. When both the initializing field and the assisting field are inverted, the Hall voltage signal originating from the initial magnetization and magnetization dynamics is inverted, while the background from the longitudinal resistance remains unchanged. Therefore, this anti-symmetrization method is effective.

Here, we show the data before and after anti-symmetrization for the cases of pulse (Fig. S3) and AC (Fig. S4). In the case of pulses, the probe current is small (10 μA), which suppresses temperature rise, allowing behaviors such as the sign change in the Hall resistance associated with magnetization reversal to be observed even before anti-symmetrization. On the other hand, in the case of AC, the background due to longitudinal resistance is significant relative to the Hall voltage signal, the anti-symmetrization procedure is necessary for the magnetization reversal behavior to be observed. In both cases, however, the behavior is the same: when the current flows in the $+x$ direction, the magnetization points in the $-z$ direction, and when the current flows in the $-x$ direction, the magnetization points in the $+z$ direction.

**Supplementary Note 2: Frequency dependence of the magnetization reversal**

We mainly use $I_x(t) = I_0 \sin(2\pi f t)$ with $I_0 = 300$ µA and $f$=10 Hz as the input current in the main text. Here we show the response waveforms $V_y(t)$ and $I_x - V_y$ curves and $I_x - R_{yx}$ curves for various frequencies $f = 11, 101, 1001, 10001$ Hz in Fig. S5. Overall, the qualitative behavior is basically independent of frequency for lower frequencies $f = 11, 101, 1001$ Hz. At the highest frequency $f = 10001$ Hz, the current is attenuated by the parasitic capacitance parallel to the sample, which causes a reduction in the current amplitude and a trivial phase change between current and Hall voltage. Therefore, at such a high frequency, it is not possible to correctly measure the Hall voltage as the response to the current in the time domain.

The frequency dependence of the responses presented here is attributed to the parasitic capacitance because the Hall voltage when the current is $I_x = 0$ was not $V_y = 0$ at the higher frequency. However, mechanisms other than parasitic capacitance through which the response depends on frequency, such as the inertia of the magnetization dynamics to the current, remain elusive. To explore such intrinsic frequency dependence, improvements in the equipment are necessary to enable high-frequency measurements without the influence of parasitic capacitance.

**Supplementary Note 3: Estimation of temperature increase caused by Joule heating**

We attribute the hysteretic behavior to the magnetization reversal caused by spin-orbit torque. However, there would be still another explanation that the temperature would show a hysteretic change because of Joule heating and accordingly the Hall response would reflect it. To exclude this possibility, we estimate the temperature variation during applying current using longitudinal resistivity as a thermometer. We first measured the temperature dependence of the longitudinal resistance $R_{xx}$ under the out-of-plane magnetic field $B_z = 1$ T (Fig. S6a), which is large enough to fix the magnetization along $z$ direction even under a large current excitation. The sensing current is as low as $I = 0.1$ μA, enabling us to ignore Joule heating. We then measured Hall resistance under AC with the amplitude of 300 μA and the frequency of 101 Hz as shown in the middle panel of Fig. S6b. At each time, we can estimate the sample temperature from the value of $R_{xx}$ as shown on the lower side. There is certainly no hysteresis in temperature and it does not exceed the critical temperature $T_C = 40$ K. This reinforces the idea that magnetization reversal occurs due to spin-orbit torque and that the hysteresis in the Hall voltage is caused by magnetization reversal.

**Supplementary Note 4: Nonlinear Hall effect (polynomial-type)**

Here, we consider the Hall voltage $V_y(t)$ in response to the current

$$I_x(t) = I_0 \sin(\omega t)$$

when $V_y$ is written as a power series of $I_x$;

$$V_y = R_{yx} I_x + R_{yxx} I_x^2 + R_{yxxx} I_x^3 + R_{yxxxx} I_x^4 + \cdots$$

where $R_{y\underbrace{xx\cdots x}_{n}}$ ($n = 1, 2, 3, 4, \cdots$) is coefficient of each order of $I_x$. Substituting the $I_x(t)$ into this, we obtain

$$V_y(t) = R_{yx} I_0 \sin(\omega t) + R_{yxx}(I_0 \sin(\omega t))^2 + R_{yxxx}(I_0 \sin(\omega t))^3 + R_{yxxxx}(I_0 \sin(\omega t))^4$$

$$+ \cdots$$

$$= R_{yx} I_0 \sin(\omega t) + R_{yxx} I_0^2 \frac{1 - \cos(2\omega t)}{2} + R_{yxxx} I_0^3 \left( \frac{3}{4} \sin(\omega t) - \frac{1}{4} \sin(3\omega t) \right)$$

$$+ R_{yxxxx} I_0^4 \left( \frac{3}{8} - \frac{1}{2} \cos(2\omega t) + \frac{1}{8} \cos(4\omega t) \right) + \cdots$$

$$= V_y^{(0)} + {V'}_y^{(1)} \sin(\omega t) + {V'}_y^{(2)} \cos(2\omega t) + {V'}_y^{(3)} \sin(3\omega t) + {V'}_y^{(4)} \cos(4\omega t) + \cdots,$$

where

$$V_y^{(0)} = I_0^2 R_{yxx}/2 + 3 I_0^4 R_{yxxxx}/8 + \cdots,$$

$${V'}_y^{(1)} = I_0 R_{yx} + 3 I_0^4 R_{yxxxx}/4 + \cdots, {V'}_y^{(2)} = -I_0^2 R_{yxx}/2 - I_0^4 R_{yxxxx}/2 + \cdots,$$

$${V'}_y^{(3)} = -I_0^3 R_{yxxx}/4 + \cdots, {V'}_y^{(4)} = I_0^4 R_{yxxxx}/8 + \cdots.$$

In this way, all the odd-harmonic components appear as sine functions, while all the even harmonic components appear as cosine functions in this power series, as mentioned in the main text.

**Supplementary Note 5: Frequency-mixing effect with and without threshold**

In the main text, we discuss the frequency-mixing effect using the current including 2 frequencies $f_1 = 37$ Hz and $f_2 = 125$ Hz. Here we describe how the mixed frequencies appear in the response.

First, let the current be $I_x(t) = I_1\sin(2\pi f_1 t) + I_2\sin(2\pi f_2 t)$. Then, we assume that the Hall voltage can be written as a power series of $I_x$;

$$V_y(t) = R_{yx}I_x(t) + R_{yxx}(I_x(t))^2 + R_{yxxx}(I_x(t))^3 + R_{yxxxx}(I_x(t))^4 + \cdots$$

where $R_{y\underbrace{xx\cdots x}_{n}}$ ($n = 1, 2, 3, 4, \cdots$) is coefficient of each order of $I_x(t)$. Substituting $I_x(t)$ into it, we obtain a lot of components with frequencies which are expressed as linear combinations of $f_1$ and $f_2$. For example, the second order term is written as below;

$$R_{yxx}(I_x(t))^2 = R_{yxx}(I_1\sin(2\pi f_1 t) + I_2\sin(2\pi f_2 t))^2$$

$$= R_{yxx}\left(I_1^2\frac{1 - \cos(4\pi f_1 t)}{2} + I_2^2\frac{1 - \cos(4\pi f_2 t)}{2}\right.$$

$$\left. + I_1 I_2\{\cos[2\pi(f_2 - f_1)t] - \cos[2\pi(f_1 + f_2)t]\}\right)$$

$$= R_{yxx}\frac{I_1^2 + I_2^2}{2} - \frac{R_{yxx}I_1^2}{2}\cos(4\pi f_1 t) - R_{yxx}\frac{R_{yxx}I_2^2}{2}\cos(4\pi f_2 t)$$

$$+ R_{yxx}I_1 I_2\cos[2\pi(f_2 - f_1)t] - R_{yxx}I_1 I_2\cos[2\pi(f_1 + f_2)t]$$

In this way, the sum- and difference-frequency components are derived. Other linear combinations of the form, $af_1 + bf_2$ ($a$ and $b$ being integer numbers), are also derived from higher order terms.

Here, we prove that the components with frequencies $af_1 + bf_2$ and $af_1 - bf_2$ have the same amplitude if $V_y(t)$ is written in a purely polynomial form of current without a threshold behavior. First, the components of $af_1 + bf_2$ and $af_1 - bf_2$ comes only from the terms of

$$[I_1\sin(2\pi f_1 t)]^{a+2m}[I_2\sin(2\pi f_2 t)]^{b+2n} \quad (m, n = 0, 1, 2, \cdots),$$

when $f_1$ and $f_2$ are coprime like 37 Hz and 125 Hz. Because the current can be also expressed as $I_x(t) = I_1\sin 2\pi f_1 t - I_2\sin 2\pi(-f_2)t$, this term should be equal to

$$[I_1\sin(2\pi f_1 t)]^{a+2m}[I_2\sin(2\pi f_2 t)]^{b+2n} = [I_1\sin(2\pi f_1 t)]^{a+2m}[-I_2\sin(-2\pi f_2 t)]^{b+2n}$$

$$= (-1)^b[I_1\sin(2\pi f_1 t)]^{a+2m}[I_2\sin(-2\pi f_2 t)]^{b+2n} \quad (m, n = 0, 1, 2, \cdots).$$

Thus, the coefficients $C_{a,b}$ of the term of $\cos[2\pi(af_1 + bf_2)t]$ (or $\sin[2\pi(af_1 + bf_2)t]$ and $C_{a,-b}$ of $\cos[2\pi(af_1 - bf_2)t]$ (or $\sin[2\pi(af_1 - bf_2)t]$) are necessarily connected by

$$C_{a,b} = (-1)^b C_{a,-b}$$

This shows that the amplitudes of the two components are equal, in accord with the observation as shown in the case of $B_x = 2$ T in Fig. 4d of the main text.

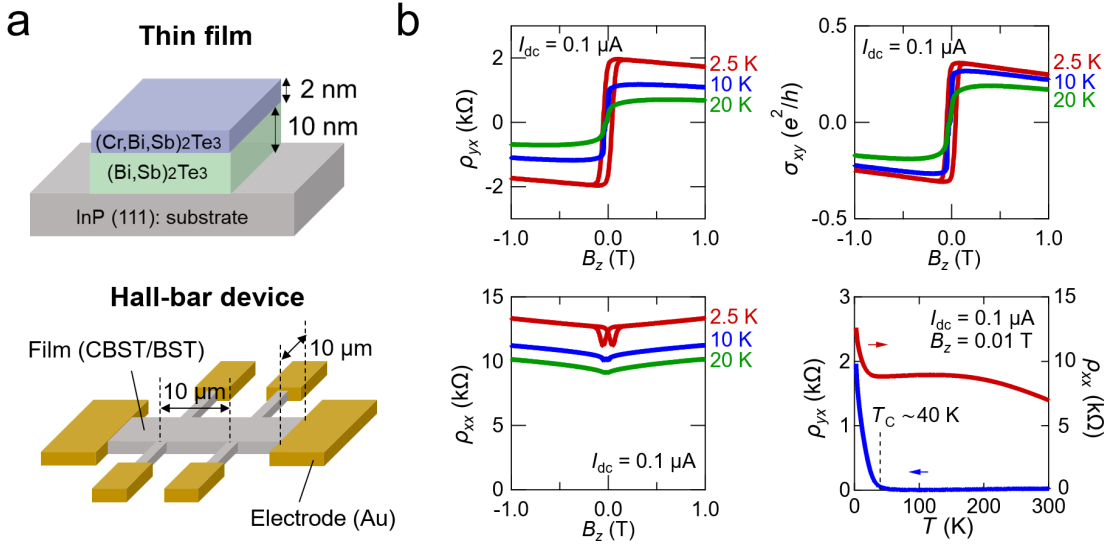

**Figure S1: Basic transport properties of the sample. a** Structure of $(Cr,Bi,Sb)_2Te_3/(Bi,Sb)_2Te_3$ thin film grown on InP(111) substrate (top) and the Hall-bar device fabricated from the film (bottom). The gray part is CBST/BST thin film and the gold parts are Au electrodes. **b** Fundamental transport properties of the sample (Hall-bar device) measured by low DC current $I_{dc} = 0.1$ μA using PPMS. The upper left, the upper right, and the lower left panels show magnetic field ($B_z$) dependence of Hall resistivity $\rho_{yx}$, Hall conductivity $\sigma_{xy}$, and longitudinal resistivity $\rho_{xx}$, respectively. All of them are measured at temperatures 2.5 K (red), 10 K (blue), 20 K (green). Temperature dependence of $\rho_{xx}$ (red) and $\rho_{yx}$ (blue) under low magnetic field $B_z = 0.01$ T are presented in the lower right panel. From the rise of the anomalous Hall resistivity ($\rho_{yx}$ at 0.01T), the transition temperature is determined as $T_C \sim 40$ K.

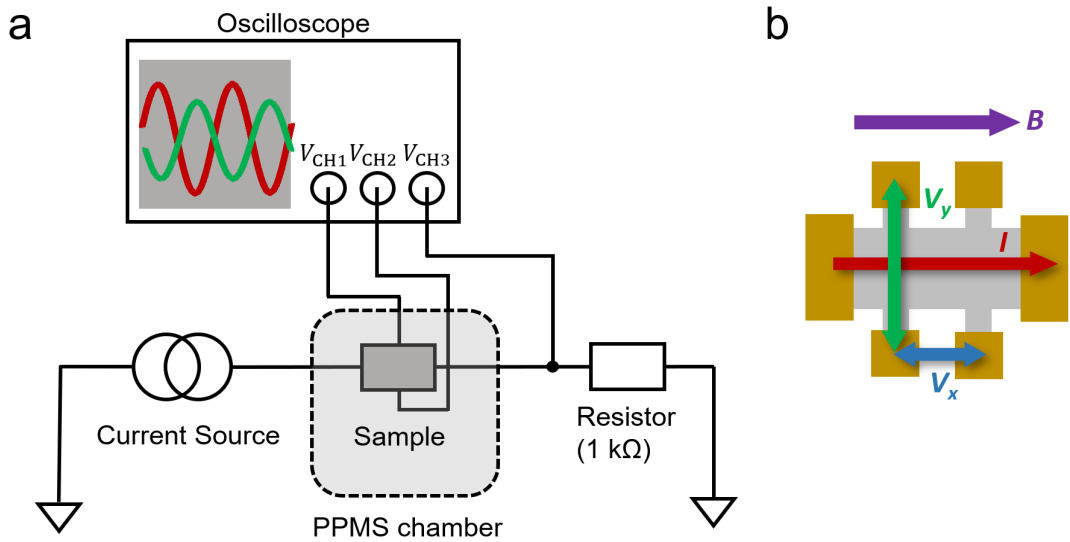

**Figure S2: Time-domain measurement setup. a** Circuit diagram of the measurement system. Current from the current source flows through the sample and resistor connected in parallel. The oscilloscope measures the Hall voltage or longitudinal voltage of the sample while simultaneously monitoring the current flowing through the system from the voltage across the resistor. The sample is put in a PPMS chamber. **b** Configuration of the current (red), the Hall voltage (green), the longitudinal voltage (blue), and the magnetic field (purple) in the sample. The shape of the sample is the same as in Fig. S1**a**.

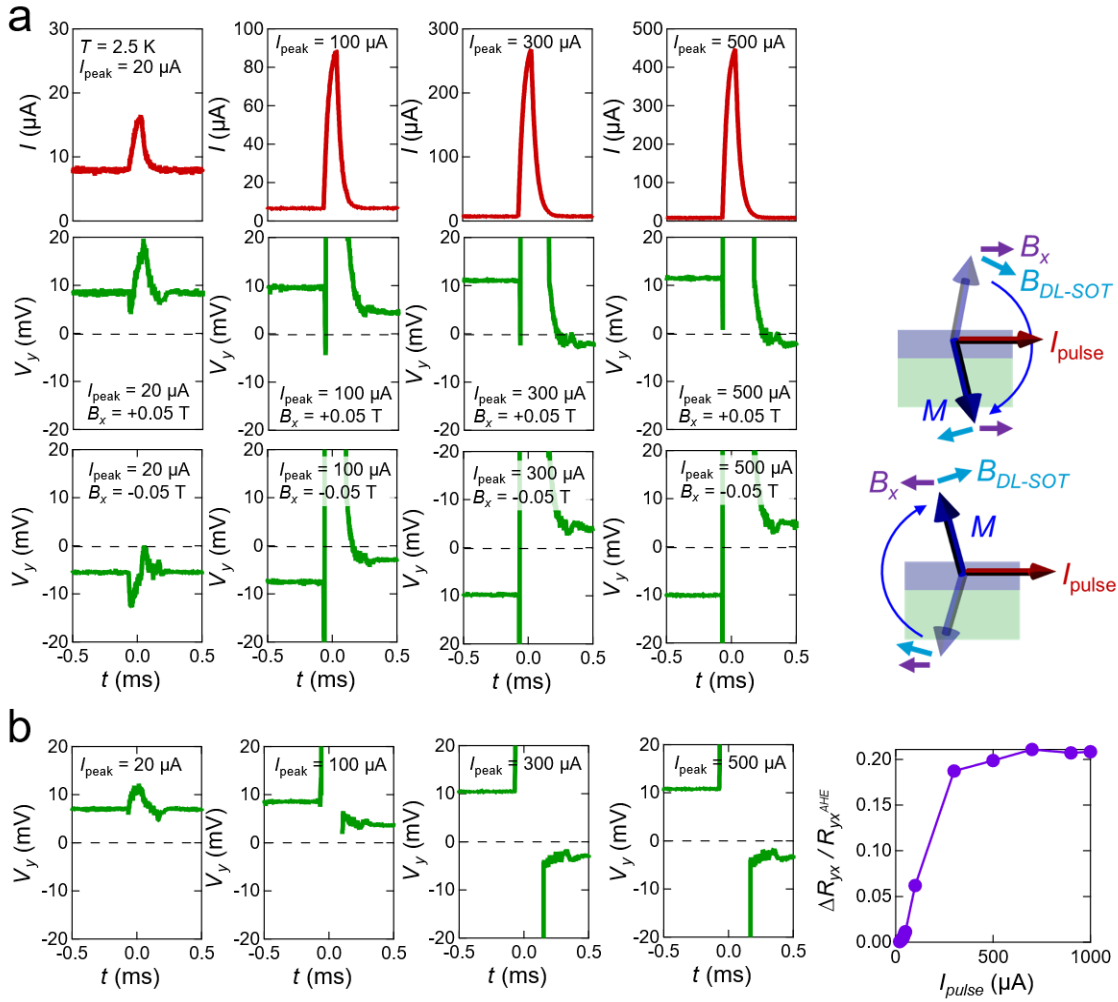

**Figure S3: Raw data and anti-symmetrized data of the time-domain measurements of magnetization reversal induced by pulse. a** Raw data of Hall voltage (green line in the middle section) in response to pulse current (red, upper) for the peak current values $I_{peak} = 20, 100, 300, 500\ \mu\text{A}$. The middle and lower sections show the responses under the in-plane magnetic field of $+0.05$ T and $-0.05$ T, respectively. The configurations of the magnetic field (purple arrow), the effective magnetic field of damping-like SOT (light blue), the magnetization (blue), and the pulse current (red) are illustrated on the right side. **b** Anti-symmetrized responses obtained from the raw data (shown in **a**) for $\pm 0.05$ T. The peak value dependence of the magnetization switching ratio is also shown in the right side. Through this figure, a different sample was used for the measurement, compared to the one used in the other figures.

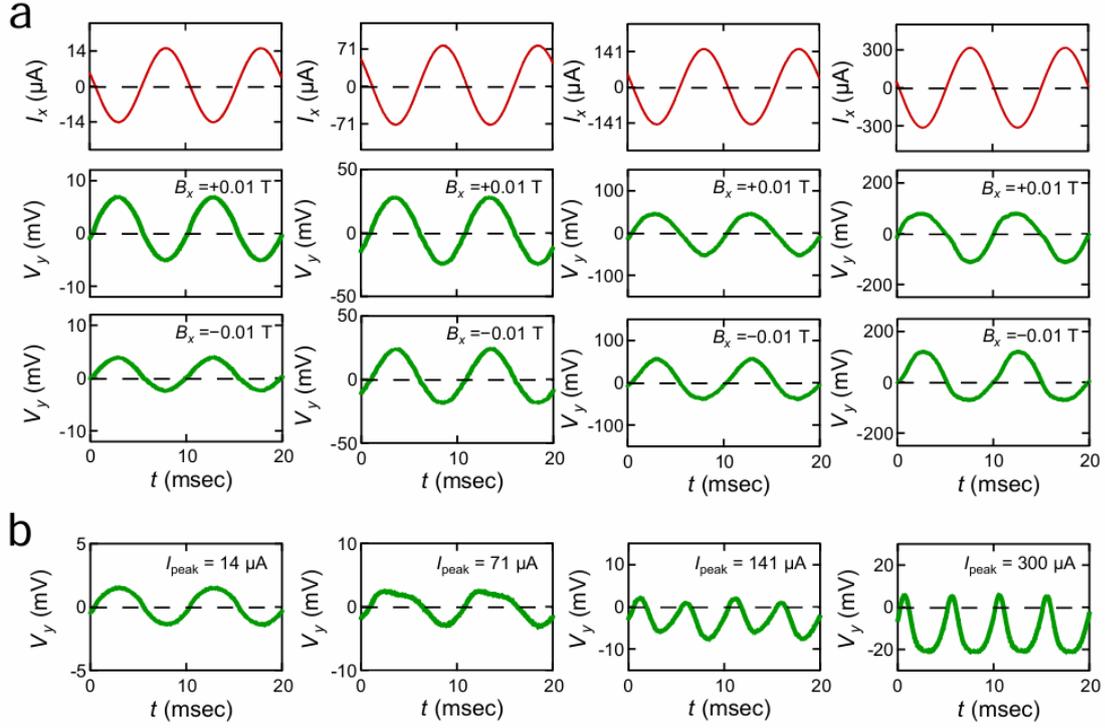

**Figure S4: Raw data and anti-symmetrized data of the time-domain measurements of magnetization reversal induced by AC. a** Raw data of Hall voltage (green lines in the middle and lower sections) in response to AC current (red, upper) for the peak current values $I_{\text{peak}} = 14, 71, 141, 300\ \mu\text{A}$. The middle and lower sections show the responses under the in-plane magnetic field of $+0.01$ T and $-0.01$ T, respectively. **b** Anti-symmetrized data obtained from the raw data (shown in **a**) for $\pm 0.01$ T. Increasing $I_{\text{peak}}$, the response $V_y(t)$ gradually becomes nonlinear, as discussed in the main text.

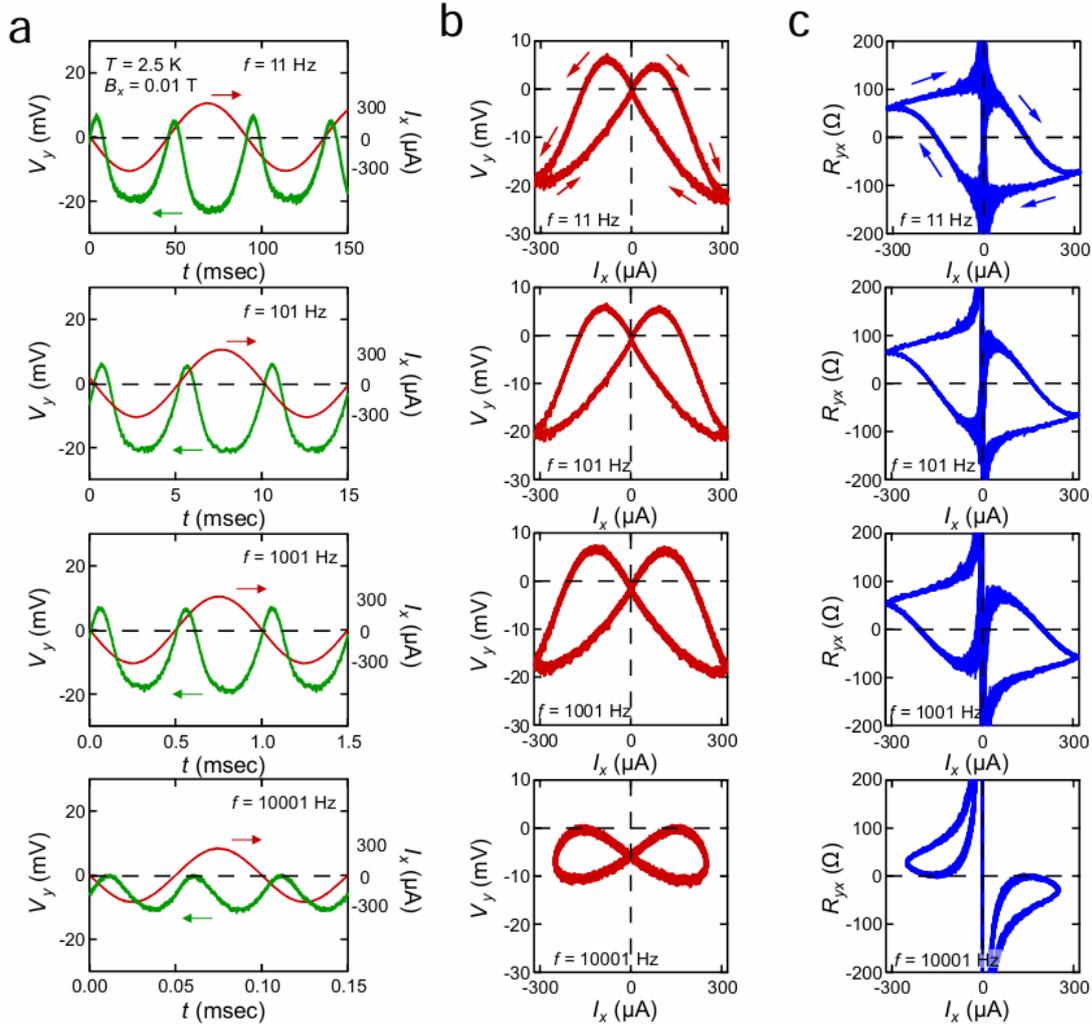

**Figure S5: Frequency dependence of AC-induced magnetization reversal. a** The Hall voltage response (green) and input current (red) with an amplitude of 300 μA and at frequencies of $11, 101, 1001, 10001$ Hz. As the frequency increases beyond 1001 Hz, the amplitude of the measured current decreases possibly due to parasitic capacitance within the electric circuit. **b** The Hall voltage vs current characteristics for each frequency. For higher frequencies 1001 Hz and 10001 Hz, the value of $V_y \neq 0$ when $I_x = 0$ because of the trivial phase rotation due to the parasitic capacitance. **c** The change in Hall resistance, calculated from the Hall voltage and current, depending on the current. For all the frequencies, $R_{yx}$ diverges around $I_x = 0$ because of $V_y/I_x = 0/0$.

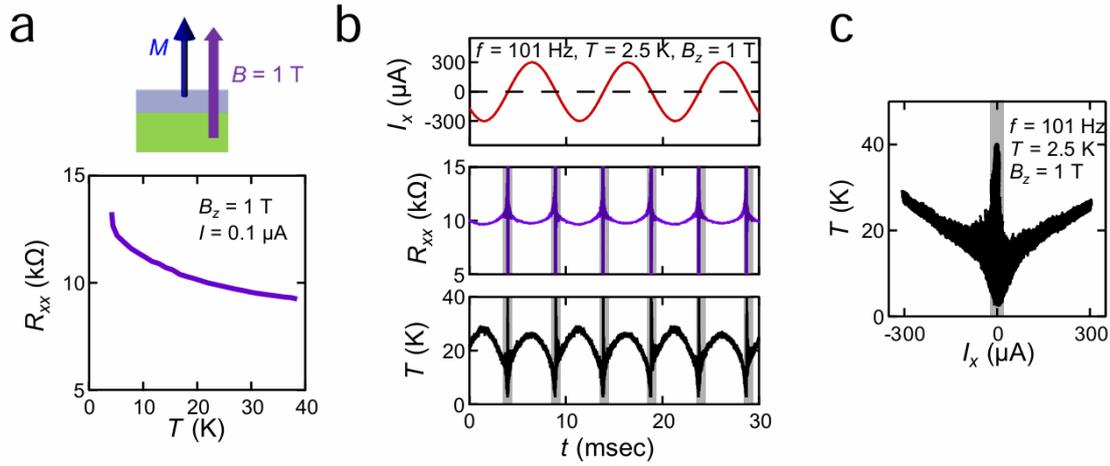

**Figure S6: Estimation of temperature increase caused by Joule heating. a** The lower part shows the temperature dependence of $R_{xx}$ measured by the probe current as low as $I = 0.1$ µA under the out-of-plane magnetization $B_z = 1$ T (the configuration is shown in the upper part). **b** Time-variation of the current $I_x$ (red), resistance $R_{xx}$ (purple), and temperature $T$ (black) which is estimated from resistance. $T$ is estimated to be as low as 30 K at the highest. $R_{xx}$ and thus $T$ diverge at $I_x = 0$ (the gray areas). **c** The dependence of $T$ on $I_x$ obtained from the $I_x(t)$ and $T(t)$ shown in **b**. $T$ increases as the absolute value of $I_x$ increases. $T$ diverges around $I_x = 0$.